\documentclass[twocolumn]{article}
\usepackage{nolta2008}
\usepackage{amsmath,amsfonts,amstext}
\usepackage{amssymb}
\usepackage{color}
\usepackage[dvips]{graphicx,graphics}
\usepackage{epsfig}
\usepackage{float}
\usepackage[english]{babel}
\usepackage[T1]{fontenc}
\usepackage[latin1]{inputenc}
\usepackage{pstricks}

\begin{document}

\title{Fractal Basins and Boundaries in 2D Maps \\ inspired in Discrete Population Models}

\author{Dani\`ele Fournier-Prunaret$^\dag$, Ricardo L\'opez-Ruiz$^\ddag$}

\address{
\dag LATTIS-INSA, LAAS-CNRS, Universit\'e de Toulouse\\
Toulouse, France\\
\ddag DIIS-BIFI, Universidad de Zaragoza \\
Zaragoza, Spain\\
Email: daniele.fournier@insa-toulouse.fr, rilopez@unizar.es
}

\maketitle  

\abstract
Two-dimensional maps can model interactions between populations. 
Despite their simplicity, these dynamical systems can show some complex situations, 
as multistability or fractal boundaries between basins that lead to remarkable pictures. 
Some of them are shown and explained here for three different 2D discrete models. \\
Keywords~: fractal, basin, two-dimensional map, fuzzy boundary.
\endabstract

\section{Introduction}

Two-dimensional maps can be used to model interactions between two different species. 
Such applications can be considered in Ecology, Biology or Economy \cite{Lo5,Cu}. 
Generally, real systems consist in a large number of interacting species 
but the understanding of the behaviour of such systems in the low dimensional case 
can be of great help as a first step attempt.
In this work, we consider two-dimensional (2D) models based on logistic 
multiplicative coupling \cite{Lo2} where complex behaviours occur such as
multistability phenomena, fractal basins of attractors and fractal boundaries 
between basins \cite{Mi1,Lo3}. These phenomena lead to some remarkable graphical representations in the phase 
space plane. In Section 2, we recall three of these considered 2D models. In Section 3, 
we emphasize these models, which permit to obtain fractal basins. Section 4 is devoted to multistability 
phenomena and fuzzy or fractal boundaries beetwen basins.

\section{The models}

\begin{figure}[htbp]
	\centering
		\includegraphics[width=0.45\textwidth]{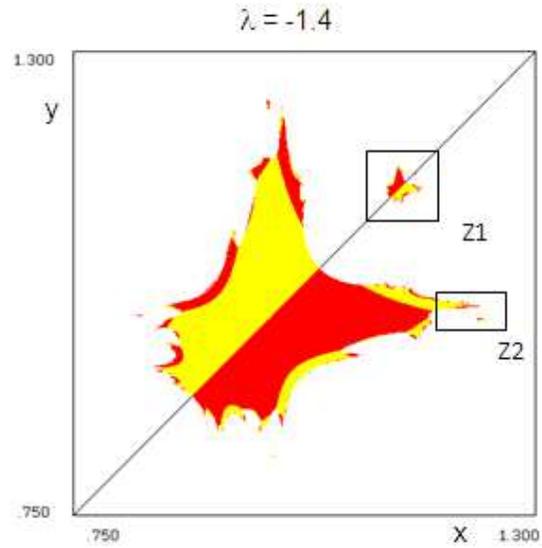}
	\caption{Fractal basins of two order 2 cycles for the map $T_1$ \eqref{1}, one basin is in yellow, 
	the other in red.}
	\label{fig:fracZ}
\end{figure}

The first considered model is the noninvertible 2D map $T_1$ defined by:
\begin{equation}\label{1}
\left\{\begin{array}{c}
x_{k+1}=\lambda (3{x_k} + 1) y_k (1-y_k) \\ 
y_{k+1}=\lambda (3{y_k} + 1) x_k (1-x_k)
\end{array}\right. 	
\end{equation}    

where $\lambda$ is a real control parameter, $x$ and $y$ are real state variables. Previous 
studies of \eqref{1} have been done in \cite{Lo3}. This model is the symmetrical case of 
a 2D model 

\begin{figure}[htbp]
	\centering
		\includegraphics[width=0.45\textwidth]{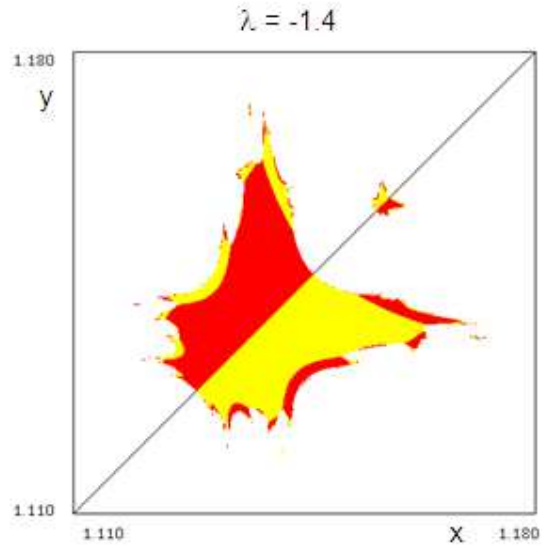}
	\caption{Enlargment of Figure~\ref{fig:fracZ}, Z1 area.}
	\label{fig:fracZ1}
\end{figure}

\begin{figure}[htbp]
	\centering
		\includegraphics[width=0.45\textwidth]{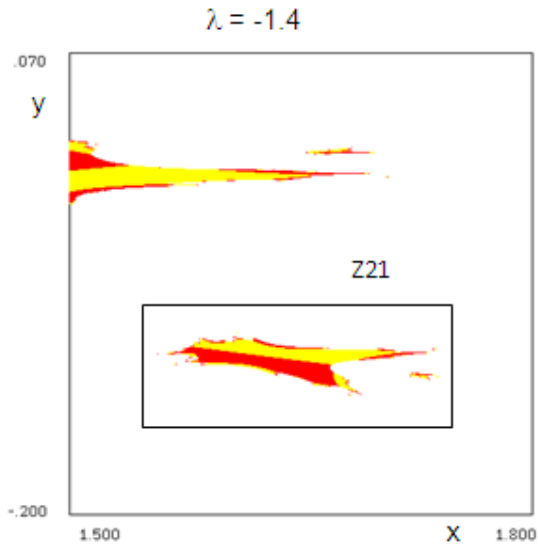}
	\caption{Enlargment of Figure~\ref{fig:fracZ}, Z2 area.}
	\label{fig:fracZ2}
\end{figure}

proposed for the symbiosis interaction between two species \cite{Lo6}. 
The second model is the noninvertible 2D map $T_2$ defined by:
\begin{equation}\label{2}
\left\{\begin{array}{c}
x_{k+1}=\lambda (3{y_k} + 1) x_k (1-x_k) \\ 
y_{k+1}=\lambda (3{x_{k+1}} + 1) y_k (1-y_k)
\end{array}\right. 	
\end{equation}    

where $\lambda$ is a real control parameter, $x$ and $y$ are real state variables. 
The map \eqref{2} is also inspired in the symbiosis case \cite{Lo6} by including 
a time asymmetric feedback. Previous studies of \eqref{2} have been presented in \cite{Fo}.

\begin{figure}[H]
	\centering
		\includegraphics[width=0.45\textwidth]{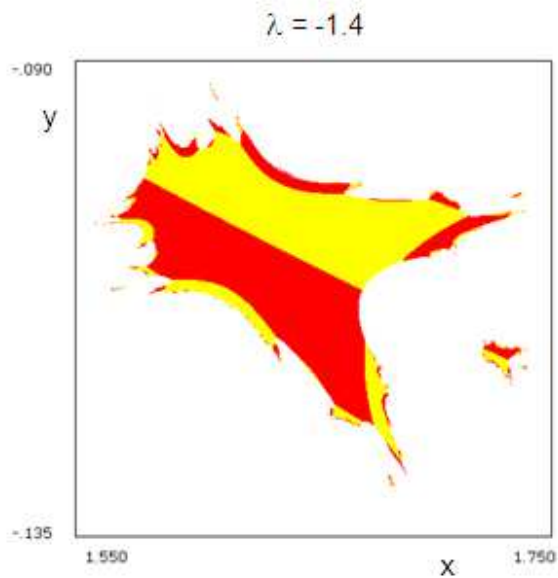}
	\caption{Enlargment of Figure~\ref{fig:fracZ2}, Z21 area.}
	\label{fig:fracZ21}
\end{figure}

The third model is the noninvertible 2D map $T_3$ defined by:
\begin{equation}\label{3}
\left\{\begin{array}{c}
x_{k+1}=\lambda (-3{y_k} + 4) x_k (1-x_k) \\ 
y_{k+1}=\lambda (-3{x_{k}} + 4) y_k (1-y_k)
\end{array}\right. 	
\end{equation} 

where $\lambda$ is real, $x$ and $y$ are real state variables. The map \eqref{3} corresponds 
to a competitive interaction between two species \cite{Lo4}. 
In each model, $\lambda$ measures the strength of the coupling. 

\begin{figure}[htbp]
	\centering
		\includegraphics[width=0.45\textwidth]{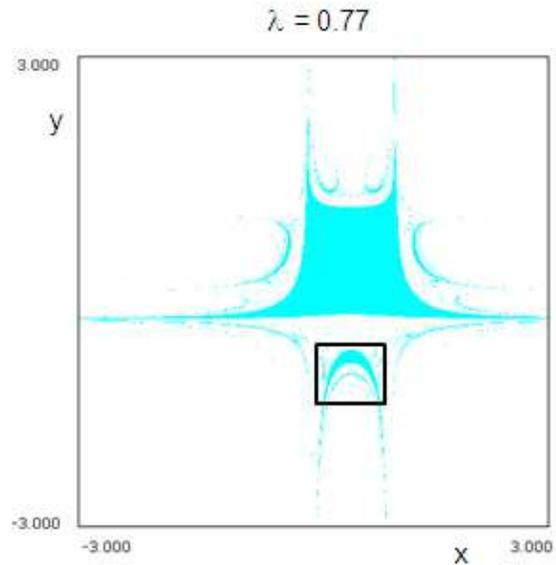}
	\caption{Basin of a fixed point for the map $T_2$ \eqref{2}.}
	\label{fig:bas1}
\end{figure}

\begin{figure}[htbp]
	\centering
		\includegraphics[width=0.45\textwidth]{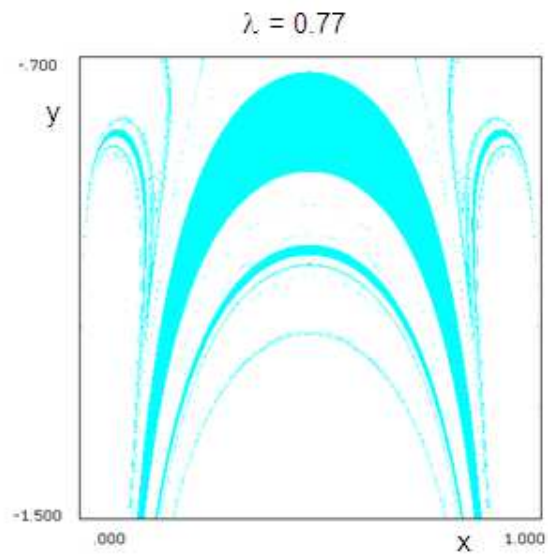}
	\caption{Enlargment of Figure~\ref{fig:bas1}. Fractal structure is observed.}
	\label{fig:bas11}
\end{figure}

\section{Fractal basins}
Figures~\ref{fig:fracZ}-\ref{fig:fracZ21} show the basin of two coexisting attractors 
and successive enlargements in the case of the map $T_1$, each attractor being an order 2 cycle: 
one basin is in yellow, the other one in red. Basins are symmetrical, non-connected and fractal. 
Successive enlargements of the 2-dimensional phase space on the diagonal (Z1 area) and on Z2 area 
show auto-similarity properties. The shape of each enlarged piece of basin is similar to the precedent
piece with an alternation between the successive locations of symmetrical yellow and red basins. 
Such a shape can be explained by using the critical manifolds of \eqref{1} \cite{Mi1,Lo3}.

Figures~\ref{fig:bas1}-\ref{fig:bas11} show the fractal basin of a fixed point for the map $T_2$. 
This basin is non-connected. Figure~\ref{fig:bas11} shows clearly the auto-similarity property. 
The appearance of such a fractal basin can also be explained by using the critical manifolds \cite{Mi1,Fo}.

\begin{figure}[H]
	\centering
		\includegraphics[width=0.45\textwidth]{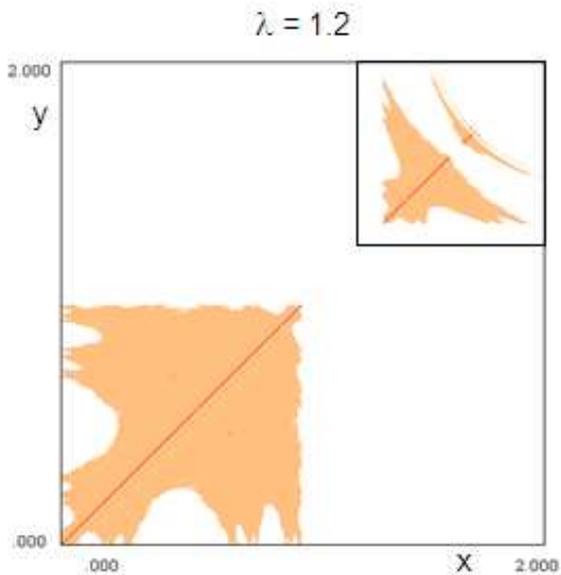}
	\caption{Fractal basin for the map $T_3$ \eqref{3}.}
	\label{fig:fraccompet}
\end{figure}

\begin{figure}[H]
	\centering
		\includegraphics[width=0.45\textwidth]{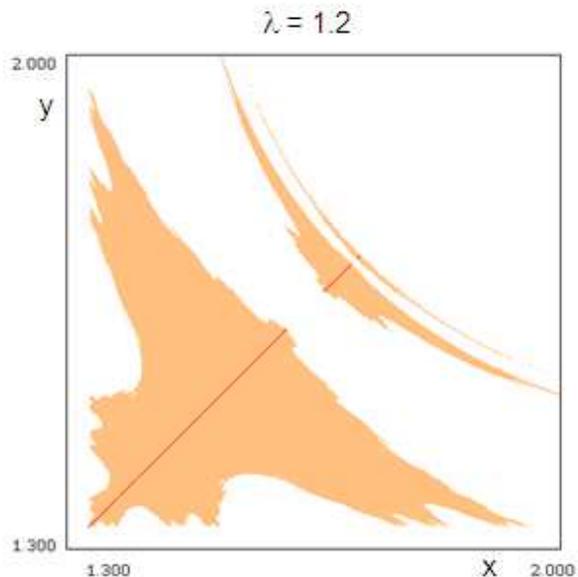}
	\caption{Enlargment of Figure~\ref{fig:fraccompet}.}
	\label{fig:fraccompetZ}
\end{figure}

Figures~\ref{fig:fraccompet}-\ref{fig:fraccompetZ} show a fractal basin for the map $T_3$. 
The attractor is chaotic and it is not represented on the Figures \cite{Lo4}. As in the case 
of the map \eqref{1}, the successive auto-similar pieces of the basin are located on the diagonal. 
The fractalization of the basins can be understood by means of the critical manifolds \cite{Mi1,Lo4}, 
as in the case of the maps \eqref{1} and \eqref{2}.

\section{Multistability and Fractal boundaries}
  
Multistability among several attractors is very frequent in these models. As it can be seen in 
the Figures~\ref{fig:bas2_3c}-\ref{fig:ridcompetZ}, the basins can take complex and fractal 
forms. Figure~\ref{fig:bas2_3c} shows the shape of the basins of three different attractors. 
The boundary among those basins is fractal, due to the accumulation and the entanglement close 
to the boundary of the square $[0,1]^2$. Figure~\ref{fig:basICC} represents the basin of two 
different attractors, wich are order 3 invariant closed curve (ICC), for the map $T_2$. Then
we obtain six different basins in $T_2^3$ for each piece of the two co-existing order 3 ICC in $T_2$.
Some of the basins can be riddled, that is, successive zooms of a basin zone cannot differentiate  
the boundaries between the basins of the different attractors. Such basins are shown on 
Figures~\ref{fig:ridcompet}-\ref{fig:ridcompetZ}. There are two chaotic attractors, which 
are order 52 chaotic rings. They are called weakly chaotic because of the greater Lyapunov exponent, 
which is slightly positive \cite{Fo}.

\end{document}